\documentclass[11pt,reqno]{amsart}
\usepackage[top=1in, left=1in, right=1in, bottom=1in]{geometry}
\usepackage[parfill]{parskip}    
\usepackage{hyperref}
\usepackage{url}
\usepackage{algorithm}
\usepackage[noend]{algorithmic}
\usepackage{amssymb, amsmath}
\usepackage{graphicx}
\usepackage{comment}
\usepackage{enumitem}
\usepackage{subcaption}
\usepackage{caption}
\usepackage{booktabs}
\usepackage{threeparttable}
\usepackage{amsaddr}

\newcommand{\RR}{\ensuremath{\mathbb{R}}}
\newcommand{\T}{\ensuremath{\top}}

\renewcommand{\leq}{\leqslant}
\renewcommand{\geq}{\geqslant}

\newcommand{\node}{\texttt{n}}
\newcommand{\parent}{\texttt{par}}
\newcommand{\anc}{\texttt{anc}}
\newcommand{\ch}{\texttt{ch}}
\newcommand{\PUBD}{\text{PUBD}}
\newcommand{\RLBD}{\text{RLBD}}
\newcommand{\A}{\mathcal{A}}
\newcommand{\Ap}{\mathcal{A}^\prime}

\newtheorem{theorem}{Theorem}[section]

\title[GOP for SMMMF]{A global optimization algorithm for sparse mixed membership matrix factorization}
\author[F. Zhang, et al]{Fan Zhang\,$^{1}$, Chuangqi Wang\,$^{1}$, Andrew C. Trapp\,$^{1}$, Patrick Flaherty\,$^{1,2}$}
\address{$^{1}$Department of Biomedical Engineering, Worcester Polytechnic Institute, MA, USA\\
	$^{2}$Department of Mathematics and Statistics, University of Massachusetts, Amherst, MA, USA}

\begin{document}
	
\maketitle

\begin{abstract}

Mixed membership factorization is a popular approach for analyzing data sets that have within-sample heterogeneity.
In recent years, several algorithms have been developed for mixed membership matrix factorization, but they only guarantee estimates from a local optimum.
Here, we derive a global optimization (GOP) algorithm that provides a guaranteed $\epsilon$-global optimum for a sparse mixed membership matrix factorization problem.
We test the algorithm on simulated data and find the algorithm always bounds the global optimum across random initializations and explores multiple modes efficiently.
\end{abstract}

\section{Introduction}

Mixed membership matrix factorization has been used in document topic modeling \cite{Blei2003a}, collaborative filtering \cite{Mackey2010}, population genetics \cite{Pritchard2000}, and social network analysis \cite{Airoldi:2008wi}. 
The underlying assumption is that an observed feature for a given sample is a mixture of shared, underlying groups. 
These groups are called topics in document modeling, subpopulations in population genetics, and communities in social network analysis. 
In bioinformatics applications the groups are called subtypes and we adopt that terminology here. 
Mixed membership matrix factorization simultaneously identifies both the underlying subtypes and the distribution over those subtypes for each individual sample.

%

\subsection{Mixed Membership Model}

The mixed membership matrix factorization problem can equivalently be viewed as
inference in a particular statistical model~\cite{Singh2008}.
These models typically have a latent Dirichlet random variable that allows each sample to have its own distribution over subtypes and a latent variable for the feature weights that describe each subtype. 
The inferential goal is to estimate the joint posterior distribution over these latent variables and thus obtain the distribution over subtypes for each sample and the feature vector for each subtype.
Non-negative matrix factorization techniques have been used in image analysis and collaborative filtering applications~\cite{Lee1999,Mackey2010}. 
Topic models for document clustering have also been cast as a matrix factorization problem~\cite{Xu2003}.

The basic mixed membership model structure has been extended in various interesting ways. 
A hierarchical Dirichlet prior allows one to obtain a posterior distribution over the number of subtypes~\cite{Teh2005}.
A prior on the subtype variables allows one to impose specific sparsity constraints on the subtypes~\cite{Kaban:2007gz,MacKay:1992ul,Taddy:2012dd}. 
Correlated information may be incorporated to improve the coherence of the subtypes~\cite{Blei2006}.

Sampling or variational inference methods are commonly used to estimate the
posterior distribution of interest for mixed membership models, but these only
provide local or approximate estimates. 
A mean-field variational algorithm~\cite{Blei2003a} and a collapsed Gibbs sampling algorithm have been developed for Latent Dirichlet Allocation~\cite{Xiao2010}. 
However, Gibbs sampling is approximate for finite chain lengths and variational inference is only guaranteed to converge to a local optimum.

\subsection{Benders' Decomposition and Global Optimization (GOP)}

In many applications it is important to obtain a globally optimal solution
rather than a local or approximate solution.
Recently, there have been significant advances in deterministic optimization methods for general biconvex optimization problems~\cite{Floudas2008, horst2013global}. 
Here, we show that mixed membership matrix factorization can be cast as a biconvex optimization problem and the $\epsilon$-global optimum can be obtained by these deterministic optimization methods.






Benders' decomposition exploits the idea that in a given optimization problem there are often ``complicating variables'' -- variables that when held fixed yield a much simpler problem over the remaining variables~\cite{Benders1962}.
Benders developed a cutting plane method for solving mixed integer optimization problems that can be so decomposed.
Geoffrion later extended Benders' decomposition to situations where the primal problem (parametrized by fixed complicating variable values) no longer needs to be a linear program~\cite{Geoffrion1972}.
The Global Optimization (GOP) approach is an adaptation of the original Benders' decomposition that can handle a more general class of problems that includes mixed-integer biconvex optimization problems~\cite{floudas2013deterministic}.
Here, we exploit the GOP approach for solving a particular mixed membership matrix factorization problem.

\subsection{Contributions}

Our contribution is bringing the GOP algorithm into contact with the mixed membership matrix factorization problem, computational improvements to the branch-and-bound GOP algorithm, and experimental results.
Our discussion of the GOP algorithm here is necessarily brief.
The details of problem conditions, convergence properties, and a full outline of the algorithm steps for the branch-and-bound version of the algorithm are found elsewhere~\cite{floudas2013deterministic}.

We outline the general sparse mixed membership matrix factorization problem in Section~\ref{sec:problem_formulation}. 
In Section~\ref{sec:algorithm}, we use GOP to obtain an $\epsilon$-global optimum solution for the mixed membership matrix factorization problem. 
In Section~\ref{sec:experiments}, we show empirical accuracy and convergence time results on a synthetic data set.
Finally, we discuss further computational and statistical issues in Section~\ref{sec:discussion}.

\section{Problem Formulation}
\label{sec:problem_formulation}

The problem data is a matrix $y \in \RR^{M \times N}$, where an element $y_{ji}$ is an observation of feature $j$ in sample $i$. 
We would like to represent each sample as a convex combination of $K$ subtype vectors, $y_i = x \theta_i$, where $x \in \RR^{M \times K}$ is a matrix of $K$ subtype vectors and $\theta_i$ is the mixing proportion of each subtype.
We would like $x$ to be sparse because doing so makes interpreting the subtypes easier and often $x$ is believed to be sparse \textit{a priori} for many interesting problems. 
In the specific case of cancer subtyping, $y_{ji}$ may be a normalized gene expression measurement for gene $j$ in sample $i$. 
We write this matrix factorization problem as

\begin{flalign}\label{eqn:opt1}
	\underset{\theta,x}{\text{minimize}}  &\  \|y_i-x\theta_i\|^2_2 \nonumber\\
	\text{subject to} &\  \|x\|_1 \leq P \\
	&\  \theta_i \in \Delta^{K-1}\  \forall i, \nonumber
\end{flalign}
where $\Delta^{K-1}$ is a $K$-dimensional simplex.

Optimization problem~\eqref{eqn:opt1} can be recast with a biconvex objective and a biconvex domain as
\begin{flalign}\label{eqn:opt2}
	\underset{\theta,x,z}{\text{minimize}}      &\  	\|y-x\theta\|^2_2 \nonumber\\
	\text{subject to}   &\  \sum_{j=1}^M \sum_{k=1}^K z_{jk} \leq P \\
                    &\  -z_{jk} \leq x_{jk} \leq z_{jk} \ \forall (j,k) \nonumber\\
                    &\  \theta_i \in \Delta^{K-1}\  \forall i, z_{jk} \geq 0\  \forall (j,k).\nonumber
\end{flalign}

If either $x$ or $\theta$ is fixed then \eqref{eqn:opt2} reduces to a convex optimization problem. 
Indeed, if $x$ is fixed, the optimization problem is a form of constrained linear regression. 
If $\theta$ is fixed, we have a form of LASSO regression.
We prove that \eqref{eqn:opt1} is a biconvex problem in Appendix \ref{bioconvex_proof}.
Since both problems are computationally simple, we could take either $x$ or $\theta$ to be the ``complicating variables'' in Benders' decomposition and we choose $\theta$.

A common approach for solving an optimization problem with a nonconvex objective function is to alternate between fixing one variable and optimizing over the other. 
However, this approach only provides a local optimum~\cite{Gorski2007a}.
A key to the GOP algorithm is the Benders'-based idea that feasibility and optimality information is shared between the primal problems in the form of constraints.

\section{Algorithm} 
\label{sec:algorithm}


We use the global optimization approach to solve for $\epsilon$-global optimum values of $x$ and $\theta$~\cite{Floudas1990,Floudas1994,floudas2013deterministic}. 
First, we partition the optimization problem decision variables into ``complicating'' and ``non-complicating'' variables.
Then, the GOP algorithm alternates between solving a \textit{primal problem} over $\theta$ for fixed $x$, and solving a \textit{relaxed dual problem} over $x$ for fixed $\theta$.
The primal problem provides an upper bound on the original optimization problem because it contains more constraints than the original problem ($x$ is fixed).
The relaxed dual problem contains fewer constraints and forms a valid global lower bound.
The algorithm iteratively tightens the upper and lower bounds on the global optimum by alternating between the primal and relaxed dual problem and tightening the relaxation in the relaxed dual problem at each iteration.

\subsection{Initialization}
\label{sec:init}

We start by partitioning the problem into a relaxed dual problem and a primal problem.
Recall our decision that the relaxed dual problem optimizes over $x$ for fixed values of the complicating variables $\theta$ and the primal problem optimizes over $\theta$.
We also initialize an iteration counter $T=1$.

At each iteration, the relaxed dual problem is solved by forming a partition of the domain of $x$ and solving a relaxed dual primal problem for each subset.
A branch-and-bound tree data structure is used to store the solution of each of these relaxed dual primal problems and we initialize the root node $\node(0)$ where $T=0$.
The parents of $\node(T)$ is denoted $\parent(\node(T))$, the set of ancestors of $\node(T)$ is denoted $\texttt{anc}(\texttt{n}(T))$, and the set of children of $\texttt{n}(T)$ is denoted $\texttt{ch}(\texttt{n}(T))$.

Finally, we initialize $x$ at a random feasible point, $x^{\node(0)}$, and store it in $\node(0)$ since we will be starting the GOP iterations by solving the primal problem over $\theta$ for a fixed $x$.

\subsection{Solve Primal Problem and Update Upper Bound}
\label{subsec:primal}

The primal problem is \eqref{eqn:opt2} constrained to fixed value of $x$ at $\node(T)$, $x^{(\node(T))}$, 

\vspace{2mm}
\fbox{
\begin{minipage}[c][10em][t]{1.0\textwidth}
\centering
\vspace{5mm}
  \textbf{Primal problem} \\
  ($x$ fixed)
    \begin{flalign*}
    \underset{\theta}{\text{minimize}}     &\  \|y-x\theta\|^2_2 \\
    \text{subject to}   &\  \theta_i^T 1_K = 1 \\
                    &\  \theta_{ki} \geq 0. 
    \end{flalign*}
\end{minipage}%
}
\vspace{2mm}

Since the primal problem is more constrained than \eqref{eqn:opt2}, the solution, $S^{(\node(T))}$, is a global upper bound.
We store the value of the upper bound, $\PUBD \leftarrow \min\{\PUBD, S^{(\node(T))} \}$, where $\PUBD$ stores the tightest upper bound.

\subsection{Solve the Relaxed Dual Problem and Update Lower Bound}
\label{subsec:dual}

The relaxed dual problem is a relaxed version of \eqref{eqn:opt2} in that it contains fewer constraints than the original problem.
Initially, at the root node $\node(0)$ the domain of the relaxed dual problem is the entire domain of $x$, $\mathcal{X}$.
Each node stores a set of linear constraints (cuts) such that when all of the constraints are satisfied, they define a region in $\mathcal{X}$.
Sibling nodes form a partition of parent's region and a node deeper in the tree defines a smaller region than shallower nodes when incorporating the constraints of the node and all of its ancestors.
These constraints are called \textit{qualifying constraints}.
Since the objective function is convex in $\theta$ for a fixed value of $x$, a Taylor series approximation of the Lagrangian with respect to $\theta$ provides a valid lower bound on the objective function.
Finally, since the objective function is convex in $\theta$, the Taylor approximation is linear and the optimal objective is at a bound of $\theta$.
The GOP algorithm as outlined in \cite{Floudas2008} makes these ideas rigorous.

The relaxed dual problem for the mixed membership matrix factorization problem \eqref{eqn:opt2} for a node $\node(T)$ is

\vspace{2mm}
\fbox{
\begin{minipage}[c][20em][t]{1.0\textwidth}
\centering
\vspace{5mm}
  \textbf{Relaxed Dual Problem} \\
  ($\theta$ fixed)
\begin{flalign*}
  \underset{Q,x,z}{\text{minimize}}     &\  Q \\
  \text{subject to}  &\  \sum_{j=1}^M \sum_{k=1}^K    z_{jk} \leq P \\ 
                           &\  -z_{jk}  \leq x_{jk} \leq z_{jk},\  z_{jk} \geq 0\\
	                   &\ \text{for} \ t \in \{ \anc(\node(T)), \node(T)\}:  \\
			   &\begin{cases}
				Q \geq L(x,\theta^{B}(t), y, \lambda^{t},\mu^{t})\big\vert^{\text{lin}}_{x^{t}, \theta^{t}} \\
				g^{t}_{ki}\big\vert^{\text{lin}}_{x^{t}}(x) \leq 0\  \text{if}\  \theta^{B}(t)_{ki} = 1\\
				g^{t}_{ki}\big\vert^{\text{lin}}_{x^{t}}(x) \geq 0\  \text{if}\  \theta^{B}(t)_{ki} = 0,
			\end{cases}
\end{flalign*}
\end{minipage}%
}
\vspace{2mm}

where $L(x,\theta^{B}(t), y, \lambda^{t},\mu^{t})\big\vert^{\text{lin}}_{x^{t}, \theta^{t}}$ is the linearized Lagrangian of \eqref{eqn:opt2}, 	$g^{t}_{ki}\big\vert^{\text{lin}}_{x^{t}}(x)$ is the $ki$-th qualifying constraint, and $\theta^B(t)$ is the value of $\theta$ at the bound such that the linearized Lagrangian is a valid lower bound in the region defined by the qualifying constraints at node $t$.
We have taken a second Taylor approximation with respect to $x$ to ensure the qualifying constraints are linear in $x$ and thus valid cuts as recommended in  \cite{Floudas2008}. 

\paragraph{Construct a child node in the branch-and-bound tree.}

A unique region in $\mathcal{X}$ for the leaf node $\texttt{ch}(\texttt{n}(T))$ is defined by the $t$-th row of $\theta^{B}$ derived from the primal problem at node $\node(T)$.
We can express this region as the qualifying constraint set,
\begin{align*}
g^{\ch(\node(T))}_{ki}\big\vert^{\text{lin}}_{x^{\node(T)}}(x) \leq 0\  \text{if}\  \theta^{B}(t)_{ki} = 1\\
g^{\ch(\node(T))}_{ki}\big\vert^{\text{lin}}_{x^{\node(T)}}(x) \geq 0\  \text{if}\  \theta^{B}(t)_{ki} = 0
\end{align*}

First, we create the $t^{th}$ child node of $\node(T)$ and populate it with this constraint set and $\theta^{B}(t)$ which will be used in the construction of the Lagrange function lower bound in the relaxed dual problem.

Second, we construct and solve the relaxed dual problem at $\ch(\node(T))$. 
First, we add the qualifying constraint sets contained in each node along the path in the branch-and-bound tree from $\ch(\node(T))$ to the root, inclusively.
For example, the qualifying constraint set for a node $\node^\prime$ along the path is
\begin{align*}
g^{\node^\prime}_{ki}\big\vert^{\text{lin}}_{x^{\node^\prime}}(x) \leq 0\  \text{if}\  \theta^{B}(\node^\prime)_{ki} = 1\\
g^{\node^\prime}_{ki}\big\vert^{\text{lin}}_{x^{\node^\prime}}(x) \geq 0\  \text{if}\  \theta^{B}(\node^\prime)_{ki} = 0,
\end{align*}
where $g^{\node^\prime}_{ki}$ is the node's $ki^{th}$ qualifying constraint, $x^{\node^\prime}$ is the node's relaxed dual problem optimizer, and $\theta^{B}(\node^\prime)$ is a $0$-$1$ vector defining the unique region for node $\node^\prime$ since $\theta_{ki} \in [0,1]$.

Third, we add the Lagrangian function lower bound constraints constructed from each node along the path in the branch-and-bound tree from $\ch(\node(T))$ to the root, inclusively.
For example the linearized Lagrange function for node $\node^\prime$,
\begin{equation*}
L(x,\theta^{B}(\node^\prime), y, \lambda^{(\node^\prime)},\mu^{(\node^\prime)})\big\vert^{\text{lin}}_{x^{(\node^\prime)}, \theta^{(\node^\prime)}}.
\end{equation*}

\paragraph{Populate the child node with the linearized Lagrange function and qualifying constraints.}
The Lagrangian function for the primal problem is 
\begin{equation}\label{eqn:lagrange1}
    \begin{split}
    L(x, \theta, \lambda, \mu) &= \sum_{i=1}^N L(x, \theta_i, \lambda_i, \mu_i)\\
    & = \sum_{i=1}^N (y_i - x \theta_i)^\T (y_i - x\theta_i) \\
    &- \lambda_i (\theta_i^\T 1_K - 1) - \mu_i^\T \theta_i\\
    &= \sum_{i=1}^N y_i^\T y_i -2 y_i^\T x \theta_i + \theta_i^\T x^\T x \theta_i \\
    & - \lambda_i (\theta_i^\T 1_K - 1) - \mu_i^\T \theta_i
    \end{split}
\end{equation}
with Lagrange multipliers $\mu \in \RR_+^{K \times N}$ and $\lambda \in \RR^N$.

The relaxed dual problem makes use of the Lagrangian function linearized about $\theta^{(t)}$ which we obtain through a Taylor series approximation,
\begin{equation}
\begin{split}
    L(x, \theta_i, \lambda_i, \mu_i) \big\vert_{\theta^{(t)}}^{\text{lin}} & \triangleq L(x, \theta_i^{(t)}, \lambda_i^{(t)}, \mu_i^{(t)}) \\
    & + \sum_{k=1}^K g_{ki}^{(t)}(x) \cdot \left(\theta_{ki} - \theta_{ki}^{(t)} \right),
\end{split}
\end{equation}
where the qualifying constraint function is
\begin{equation}\label{eqn:qc1}
\begin{split}
g^{(t)}_{i}(x) \triangleq & \nabla_{\theta_i} L \left(\theta_i, x, \lambda_i^{(t)}, \mu_i^{(t)} \right) \big\vert_{\theta_i^{(t)}} \\
& = -2 y_i^\T x + 2 \theta^{(t)\T}_i x^\T x \\
& - 1_K^\T \lambda_i^{(k)} - \mu_i^{(k)\T}.
\end{split}
\end{equation}

The qualifying constraint $g^{(t)}_{i}(x)$ is quadratic in $x$.
However, we require it to be linear in $x$ to yield a convex domain if $g^{(t)}_{i}(x) \geq 0$ or $g^{(t)}_{i}(x) \leq 0$. 
So, we linearize the Lagrangian first with respect to $x$ about $x^{(t)}$ then about $\theta_i$ at $\theta_i^{(t)}$. 
While the linearized Lagrangian is not a lower bound everywhere in $x$, it is a valid lower bound in the region bound by the qualifying constraints with $\theta_i$ set at the corresponding bounds in the Lagrangian function.

The Lagrangian function linearized about $x^{(t)}$ is
\begin{equation}
\begin{split}
L(y_i, \theta_i, x, \lambda_i, \mu_i) \bigg\vert^{\text{lin}}_{x^{(t)}} \triangleq & y_i^T y_i - \theta_i^\T x^{(t)\T} x^{(t)} \theta_i \\
& - 2 y_i^\T x \theta_i + 2 \theta_i^\T x^{(t)\T} x \theta_i \\
& - \lambda_i(\theta_i^\T 1_K - 1) - \mu_i^\T \theta_i.
\end{split}
\end{equation}

Subsequently, the Lagrangian function linearized about $(x^{(t)}, \theta_i^{(t)})$ is
\begin{equation}
\begin{split}
L(y_i, \theta_i, x, \lambda_i, \mu_i) &  \bigg|^{\text{lin}}_{x^{(t)}, \theta_i^{(t)}} \triangleq\  y_i^\T y_i + \theta_i^{(t)\T} x^{(t)\T} x^{(t)} \theta_i^{(t)} \\
& -2 \theta_i^{(t)\T} x^{(t)\T} x^{(t)} \theta_i \\
& - \lambda_i(1_K^\T \theta_i - 1) - \mu_i^\T \theta_i \\
& - 2 \theta_i^{(t)\T} x^\T x^{(t)} \theta_i^{(t)\T} - 2 y_i^\T x \theta_i \\
&  + 2 \theta_i^{(t)\T} (x^{(t)\T} x + x^\T x^{(t)}) \theta_i
\end{split},
\end{equation}
and the gradient used in the qualifying constraint is
\begin{equation}
	\begin{split}
		g^{(t)}_i\big\vert^{\text{lin}}_{x^{(t)}}(x) &\triangleq \nabla_{\theta_i} \left[ L(y_i, \theta_i, x, \lambda_i, \mu_i) \bigg|^{\text{lin}}_{x_0} \right] \bigg|_{\theta_i^{(t)}} \\
		& = -2 x^{(t)\T} x^{(t)} \theta_i^{(t)} -2 x^\T y_i \\
		& + 2 (x^{(t)\T} x + x^\T x^{(t)}) \theta_i^{(t)} - \lambda_i 1_K - \mu_i.
\end{split}
\end{equation}

\paragraph{Solve the relaxed dual problem at the child node.}
Once the valid qualifying constraints from the previous $t=1,\ldots,T-1$ iterations have been identified and incorporated, the constraint for the current $T^{th}$ iteration is
\begin{eqnarray*}
Q \geq & L(x,\theta^{B_T}, y, \lambda^{(t)},\mu^{(t)})\big\vert^{\text{lin}}_{x^{(t)}, \theta^{(t)}}\\
&g_{ki}^{(T)}\big\vert^{\text{lin}}_{x^{(t)}}(x) \leq 0 \ \text{if}\ \theta^{B_T}_{ki} = 1 \\
&g_{ki}^{(T)}\big\vert^{\text{lin}}_{x^{(t)}}(x) \geq 0 \ \text{if}\ \theta^{B_T}_{ki} = 0.
\end{eqnarray*}

The resulting relaxed dual problem is a linear program and can be solved efficiently using the off-the-shelf LP solver Gurobi~\cite{optimization2012gurobi}.
We store the optimal objective function value and the optimizing decision variables in the node.

\paragraph{Update the lower bound.} 

The global lower bound is provided by the lowest lower bound across all the leaf nodes in the branch-and-bound tree.
We store this global lower bound in a variable, RLBD.
Operationally, we maintain a dictionary where the value of a record is a pointer to a branch-and-bound tree node and the key is the optimal value of the relaxed dual problem at that leaf node.
Using this dictionary, we select the smallest key and bound to the node of the tree indicated by the value.
This element is eliminated from the dictionary since at the end of the next iteration, it will be an interior node and not available for consideration.
We increment the iteration count $T \leftarrow T+1$ and we update the global lower bound RLBD with the optimal value of the relaxed dual problem at the new node.

\paragraph{Check convergence.}
Since we always select the lowest lower bound provided by the relaxed dual problem, the lower bound is non-decreasing.
If our convergence criteria $\text{PUBD} - \text{RLBD} \leq \epsilon$ has been met, then we exit the algorithm and report the optimal $\theta$ from the node's primal problem and the optimal $x$ from the node's relaxed dual problem.
Finite $\epsilon$-convergence and $\epsilon$-global optimality proofs can be found in~\cite{Floudas1994}.

\section{Computational Improvements}
\label{sec:computational_improvements}

In the relaxed dual problem branch-and-bound tree, a leaf node below the current node $\node(T)$ is constructed for each unique region defined by the hyperplane arrangement. 
In the GOP framework, there are $KN$ hyperplanes, one of each so-called ``connected variable'' and all of the $KN$ elements of $\theta$ are connected variables.
So, an upper bound on the number of regions defined by $KN$ cuts is $2^{KN}$ because each region may be found by selecting a side of each cut.
Thus we have the computationally complex situation of needing to solve a relaxed dual problem for each of the $2^{KN}$ possible regions.

Let an arrangement $\A$ denote a set of hyperplanes and $r(\mathcal{A})$ denote the set of unique regions defined by $\A$.
In our particular situation, all of the hyperplanes pass through the unique point $x^{(\node(T))}$, so all of the regions are unbounded except by the constraints provided in $\mathcal{X}$.
A recursive algorithm for counting the number of regions $|r(\A)|$ known as Zaslavsky' Theorem, is outlined in \cite{zaslavsky1975facing}.
Indeed, $|r(\A)|$ is often much less that $2^{|\A|}$.
However, due to its recursive nature, computing the number of hyperplanes using Zaslavsky's theorem is computationally slow.

\subsection{Cell Enumeration Algorithm}
\label{subsec:cell_enumeration}

We have developed an A-star search algorithm for cell enumeration to simultaneously identify and count the set of unique regions defined by arrangement $\A$ with sign vectors.
First, we preprocess the arrangement $\A$ to eliminate trivial and redundant hyperplanes.
We eliminate a hyperplane from $\A$ if the coefficients are all zero and eliminate duplicate hyperplanes in $\A$ (see Appendix \ref{duplicate}).
We are left with a reduced arrangement, $\A^\prime$.

Here we define two concepts, \textit{strict hyperplane} and \textit{adjacent region}.
A strict hyperplane is defined as non-redundant bounding hyperplane in a single region.
If two regions exist that have sign vectors differing in only one hyperplane, then this hyperplane is a strict hyperplane.
We define an adjacent region of region $r$ as a neighbor region of $r$ if they are separated by exactly one strict hyperplane.
The general idea of the A-star algorithm uses ideas from partial order sets.
We first initialize a root region using an interior point method and then determine all of its adjacent regions by identifying the set of strict hyperplanes.
This process guarantees that we can enumerate all unique regions.

We define $\theta^{B} \in \{0,1\}^{|r(\Ap)| \times KN}$. 
The rows are regions and there are $KN$ columns. 
Each element of this matrix is either $0$ or $1$.
The $b^{th}$ region in $r(\Ap)$ is uniquely identified by the zero-one vector in the $b^{th}$ row of $\theta^{B}$.
If the $b^{th}$ element of the $ki^{th}$ row of $\theta^{B}$ is $+1$, then $g_{ki} \leq 0$.
Similarly, if the $b^{th}$ element of the $ki^{th}$ row of $\theta^{B}$ is $0$, then $g_{ki} \geq 0$.
The A-star search algorithm completes the $\theta^{B}$ matrix for the current node $\node(T)$ and a leaf node is generated for each row of $\theta^{B}$.
Thus each unique region defined by the qualifying constraint cuts provided by the Lagrange dual of the primal problem at the current node.
The details of the A-star search algorithm are covered in Section \ref{A}.


\section{Experiments}
\label{sec:experiments}

In this section, we present our experiments on synthetic data sets and show accuracy and convergence speed.
Computational complexity is evaluated by both the theoretical and empirical time complexity.

\subsection{Illustrative Example}
\label{subsec:illustrative_example}

We use a simple data set to show the operation of the algorithm in detail and facilitate visualization of the cut sets. The data set, $y$, and true
decision variable values, $(x^*, \theta^*)$, are
\begin{eqnarray*}
	& x^* =
	\left[ 
	\begin{array}{cc} 
		0, & -1
	\end{array} 
	\right],
	\theta^* = 
	\left[ 
	\begin{array}{ccc} 
		1, & 0, & 0.5 \\
		0, & 1, & 0.5
	\end{array} 
	\right],\\
	& y = 
	\left[ 
	\begin{array}{ccc} 
		0, & -1, & -0.5
	\end{array} 
	\right].
\end{eqnarray*}
\vspace{1mm}

We ran the GOP algorithm with sparsity constraint variable $P=1$ and convergence tolerance $\epsilon = 0.01$. 
There are $KN = 6$ connected variables, so we solve at most $2^{KN} = 64$ relaxed dual problems at each iteration. 
These relaxed dual problems are independent and can be distributed to different computational threads or cores. 
The primal problem is a single optimization problem and will not be distributed. 
The optimal decision variables after 72 iterations are
\vspace{5mm}
\begin{eqnarray*}
	&\hat{x} = x^{(72)} =
	\left[ 
	\begin{array}{cc} 
   	    0.080 , & -0.920
	\end{array} 
	\right],\\
	&\hat{\theta} = \theta^{(72)} =
	\left[ 
	\begin{array}{ccc} 
		1.00, & 0.080, & 0.580 \\
		0.00, & 0.920, & 0.420
	\end{array} 
	\right],
\end{eqnarray*}
\vspace{1mm}

and the Lagrange multipliers are $\hat{\lambda} = [-0.147, 0, 0]$ and $\hat{\mu} = [0, 0, 0; 0.160, 0, 0]$.

\begin{figure}[!htb]
\captionsetup[sub]{font=small}
\begin{subfigure}[b]{0.48\textwidth}
\includegraphics[width=1\textwidth]{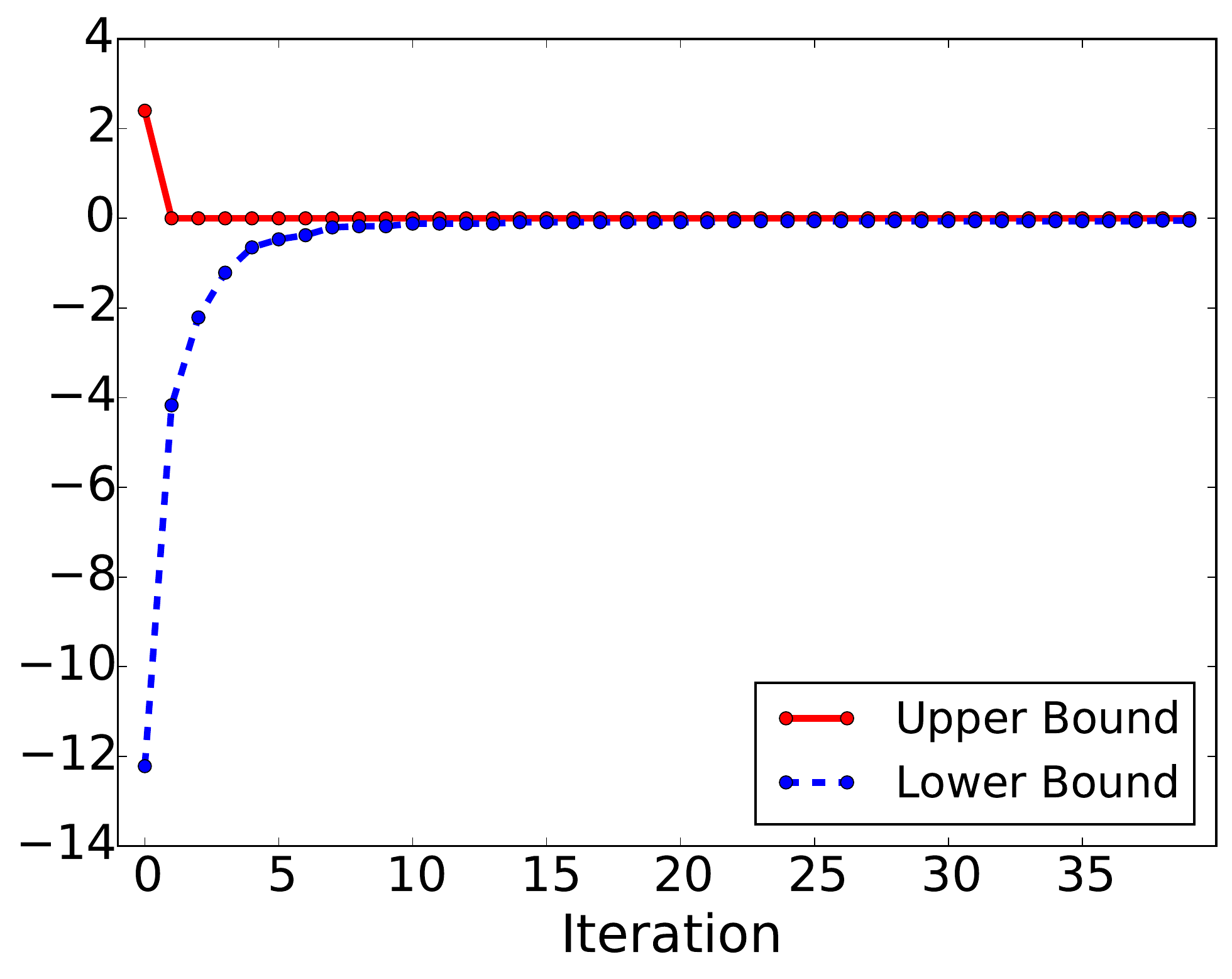}
\caption{Upper and lower bounds.}
\end{subfigure}~
\begin{subfigure}[b]{0.5\textwidth}
\includegraphics[width=1\textwidth]{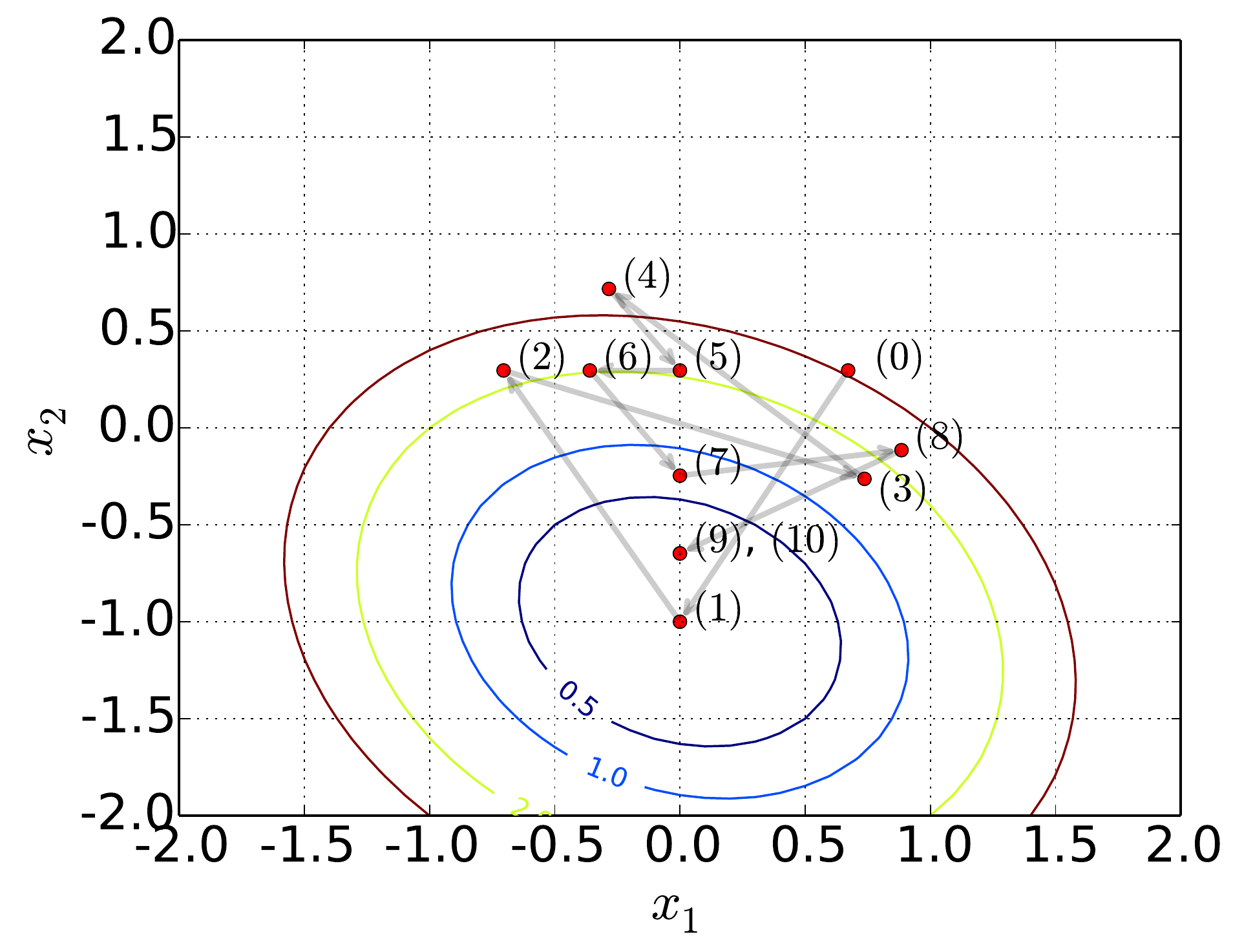}
\caption{Optimal relaxed dual problem decision variables.}
\end{subfigure}
\caption{GOP inference optimal values and optimizing $x$ variables.}
\label{fig:bounds_and_trace}
\end{figure}

Figure~\ref{fig:bounds_and_trace} (a) shows the convergence of the upper and lower bounds by iteration. 
The upper bound converges quickly and the majority of the time in the algorithm is spent proving optimality. 
With each iteration regions of the solution space are tested until the lower bound is tightened sufficiently to meet the stopping criterion. 
Figure~\ref{fig:bounds_and_trace} (b) shows the first ten $x$ values considered by the algorithm with isoclines of the objective function with $\theta^*$ fixed. 
It is evident that the algorithm is not performing hill-climbing or any other gradient ascent algorithm during its search for the global optimum.
Instead, the algorithm explores a region bound by the qualifying constraints to construct a lower bound on the objective function. 
We run it using 20 random initial values and the optimal objective functions for all random initializations are all 0, which shows that the GOP algorithm found the globally optimal solutions of this small instance.  
Furthermore, the algorithm does not search nested regions, but considers previously explored cut sets (Figure~\ref{fig:bounds_and_trace} (b)).

\begin{figure}[!htb]
\captionsetup[sub]{font=small}
\begin{subfigure}[b]{0.6\textwidth}
\includegraphics[width=1\textwidth]{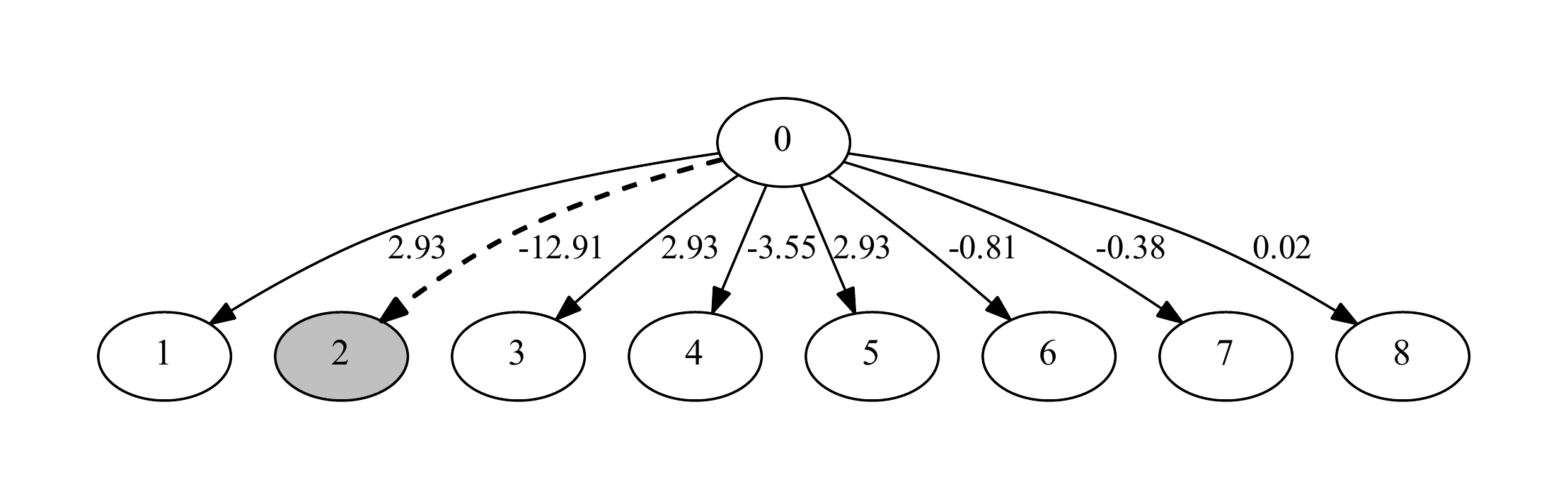}
\caption{Branch-and-bound tree at iteration 1}
\end{subfigure}~
\begin{subfigure}[b]{0.4\textwidth}
\includegraphics[width=1\textwidth]{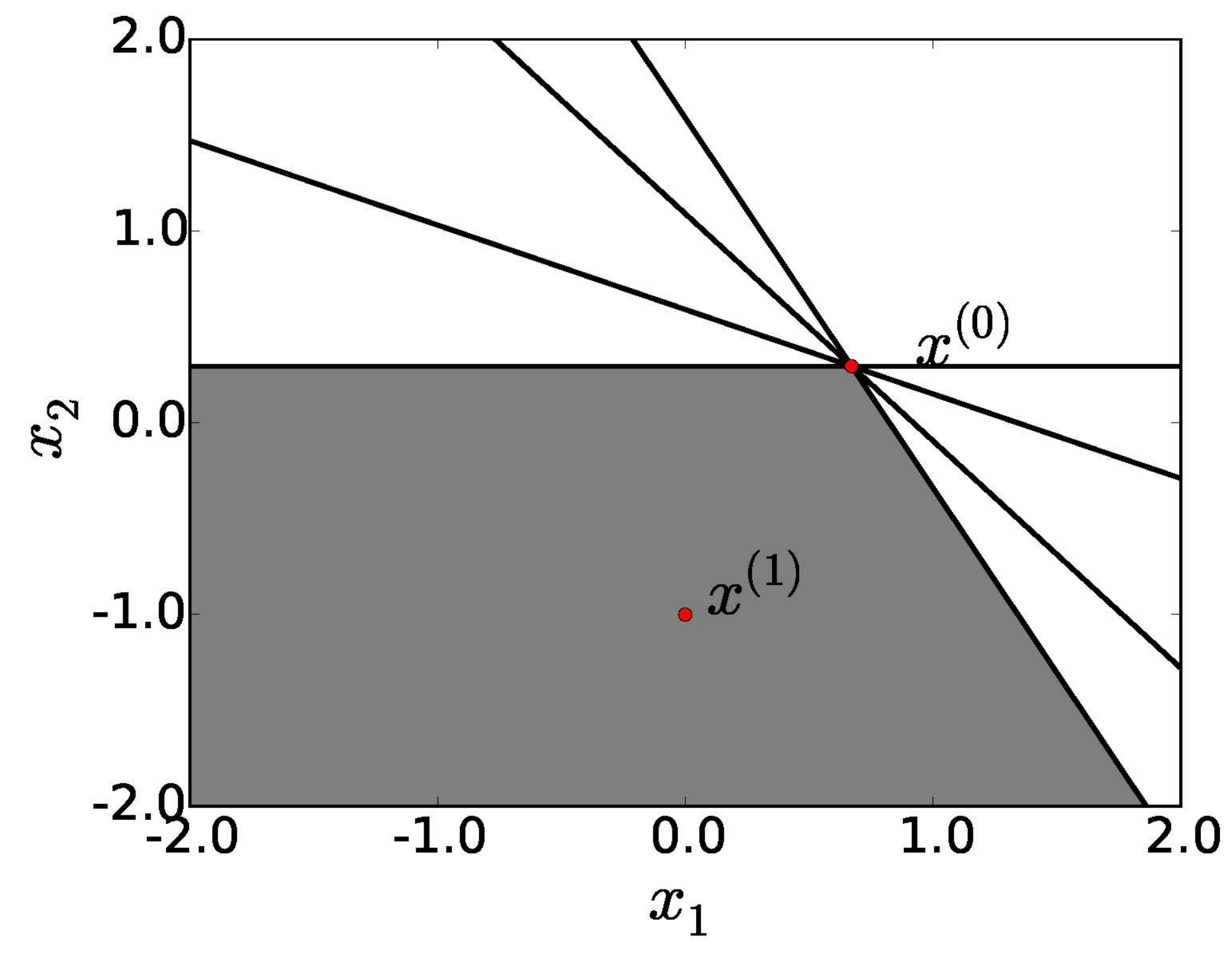}
\caption{$x$-space region at iteration 1}
\end{subfigure}

\begin{subfigure}[b]{0.6\textwidth}
\includegraphics[width=1\textwidth]{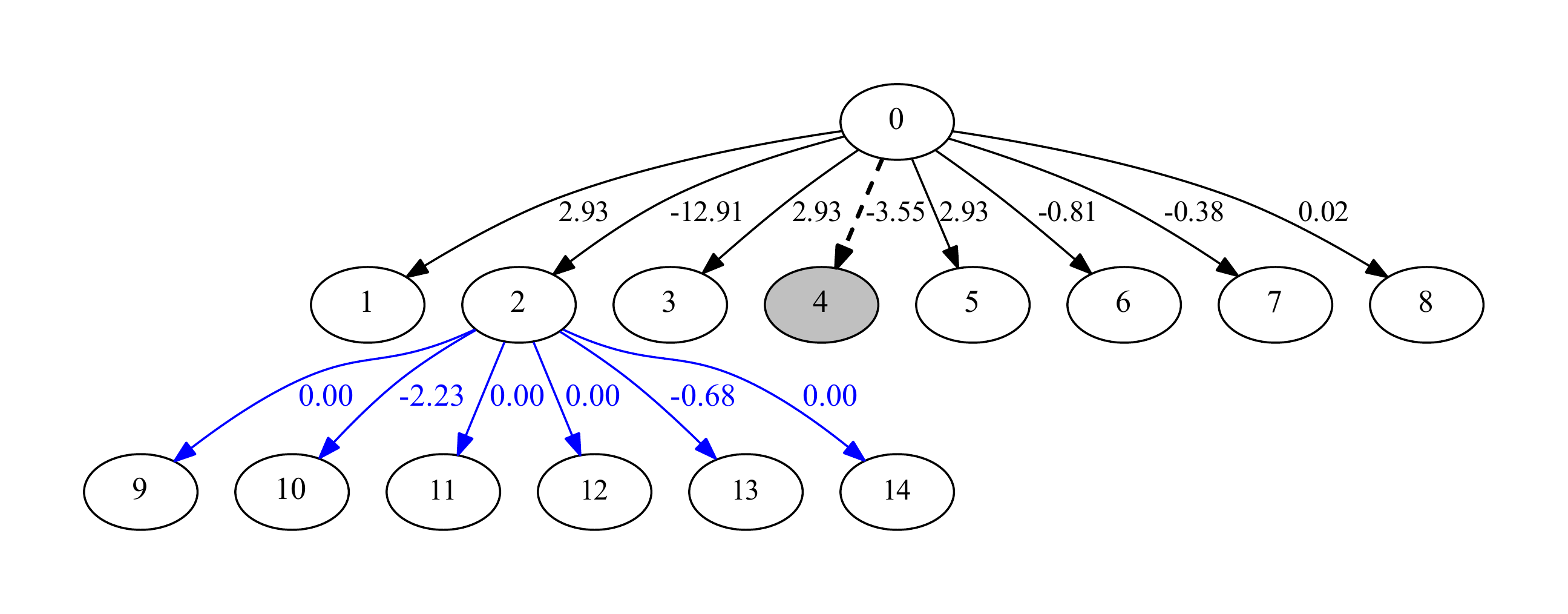}
\caption{Branch-and-bound tree at iteration 2}
\end{subfigure}~
\begin{subfigure}[b]{0.4\textwidth}
\includegraphics[width=1\textwidth]{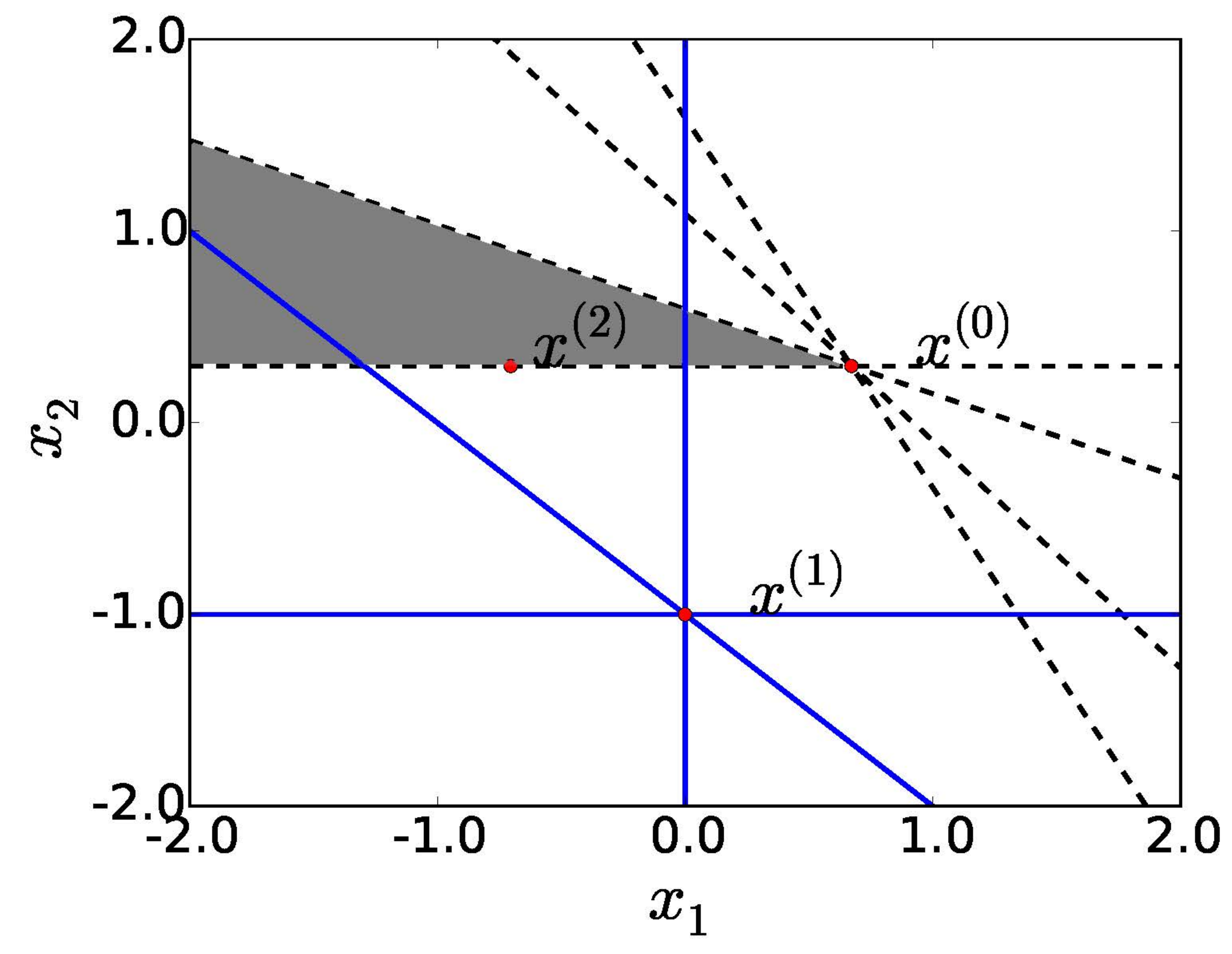}
\caption{$x$-space region at iteration 2}
\end{subfigure}

\begin{subfigure}[b]{0.6\textwidth}
\includegraphics[width=1\textwidth]{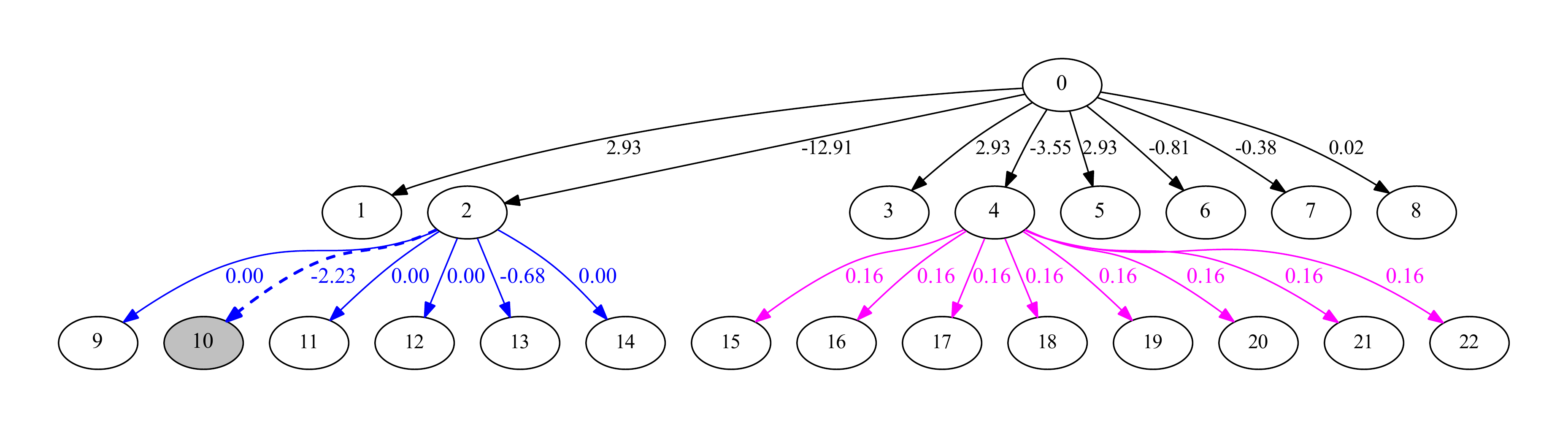}
\caption{Branch-and-bound tree at iteration 3}
\end{subfigure}~
\begin{subfigure}[b]{0.4\textwidth}
\includegraphics[width=1\textwidth]{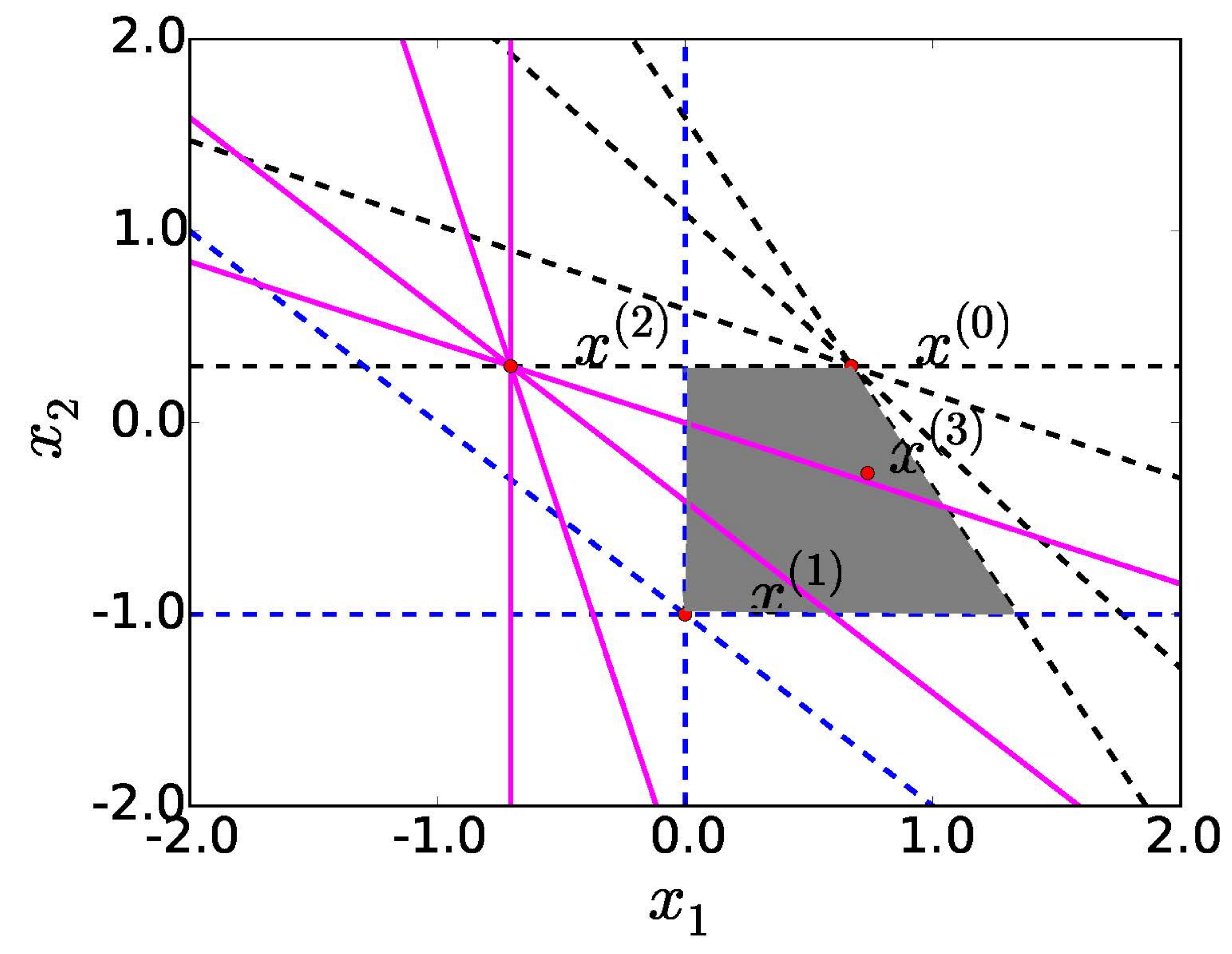}
\caption{$x$-space region at iteration 3}
\end{subfigure}
\caption{GOP branch-and-bound tree and corresponding $x$-space region. The gray node indicates the current node. The numbers on the edges indicate the optimal value of the relaxed dual problem.}
\label{fig:bb_tree}
\end{figure}

Figure~\ref{fig:bb_tree} shows the branch-and-bound tree and corresponding $x$-space region with the sequence of cut sets for the first three iterations of the algorithm.
One cut in Figure~\ref{fig:bb_tree} (b, d, f) is obtained for each of the $KN$ qualifying constraints. 
We initialize the algorithm at $x^{(0)}$.

\subsection{Accuracy and Convergence Speed}
\label{subsec:accuracy_speed}

\begin{figure}[!htb]
\captionsetup[sub]{font=footnotesize}
\begin{subfigure}[b]{0.9\textwidth}
\includegraphics[width=1\textwidth]{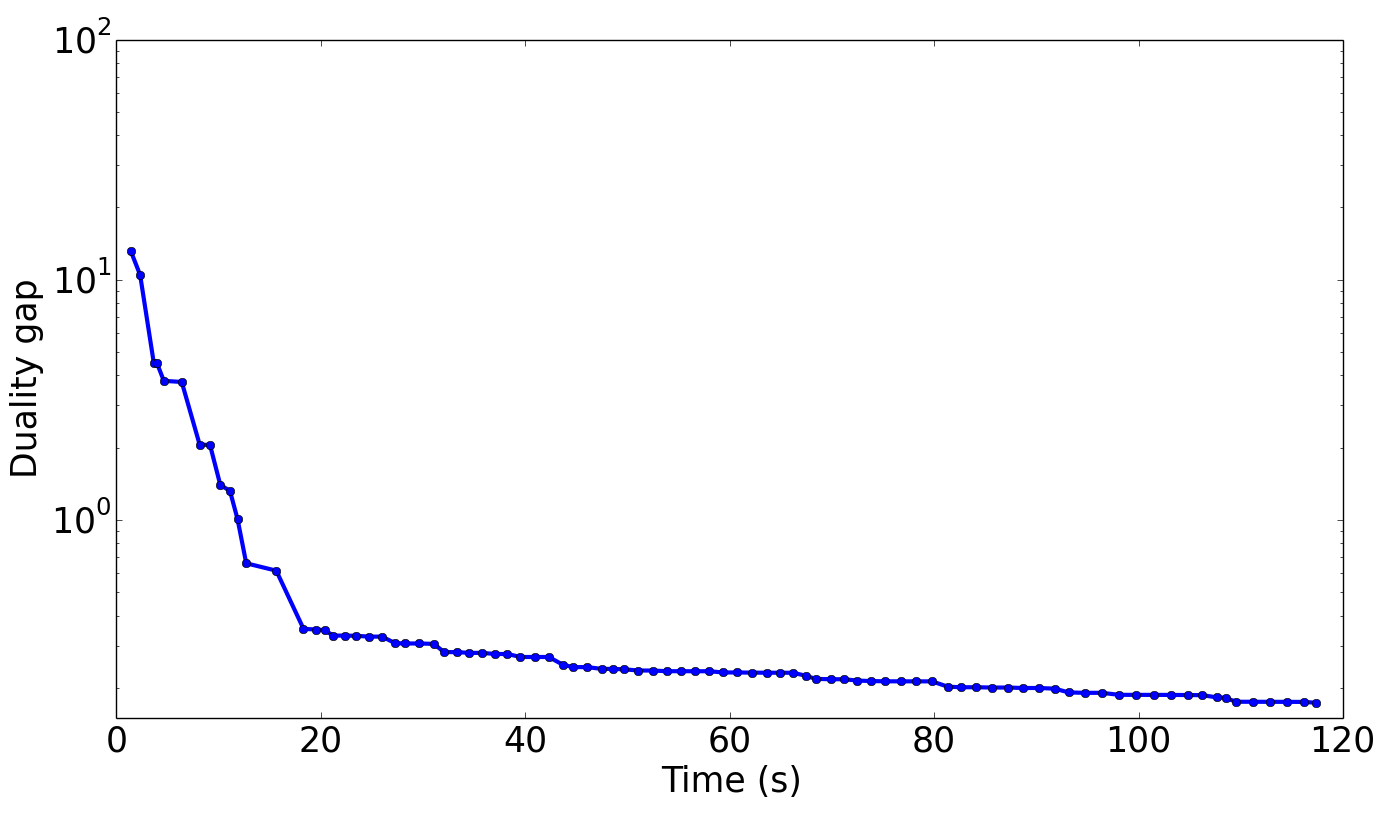}
\caption{Duality gap through the first 120 seconds.}
\end{subfigure}

\begin{subfigure}[b]{0.5\textwidth}
\includegraphics[width=1\textwidth]{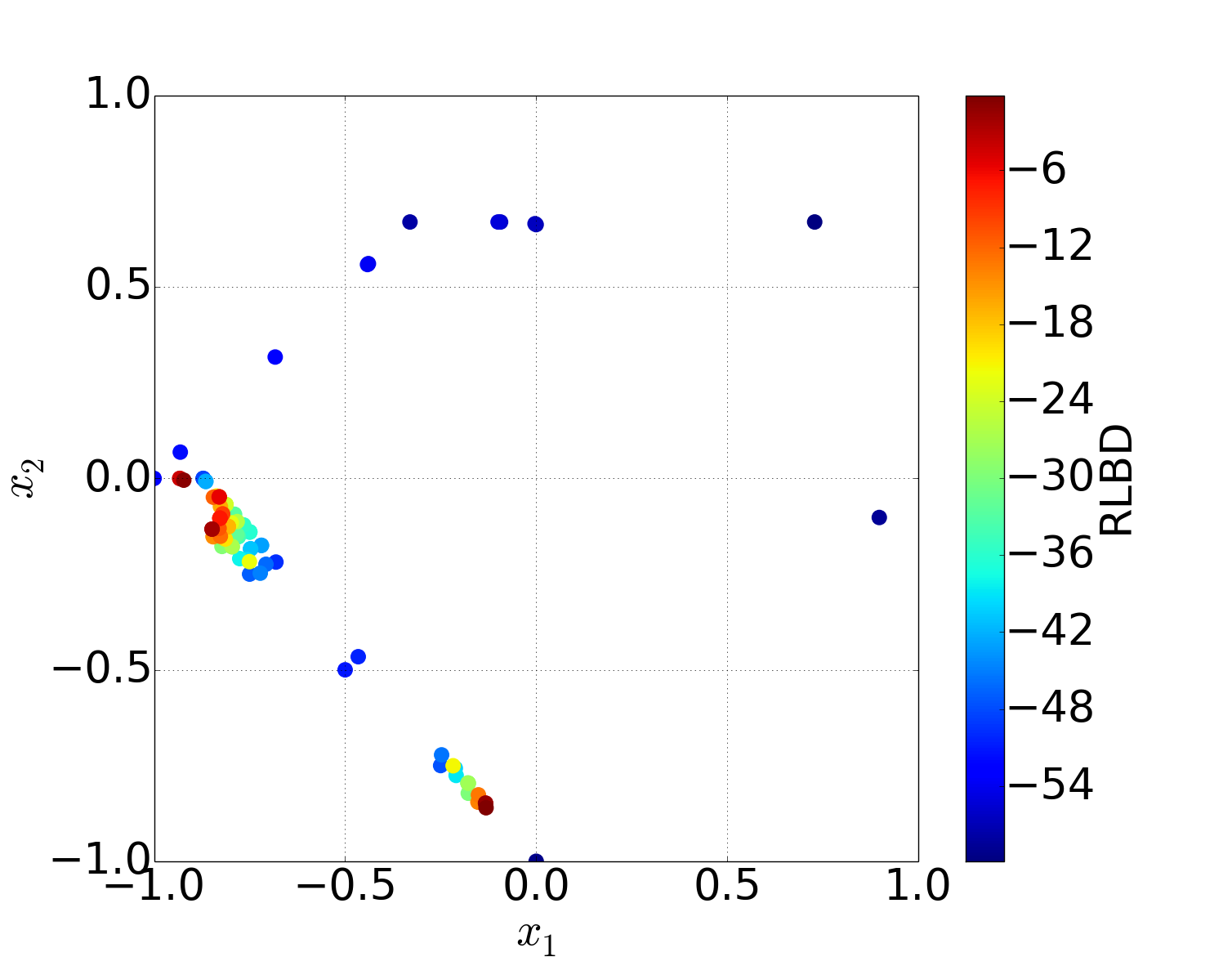}
\caption{Optimal $x$ of each iteration.
The true $x$ is (0, -1).}
\end{subfigure}~
\quad
\begin{subfigure}[b]{0.495\textwidth}
\includegraphics[width=1\textwidth]{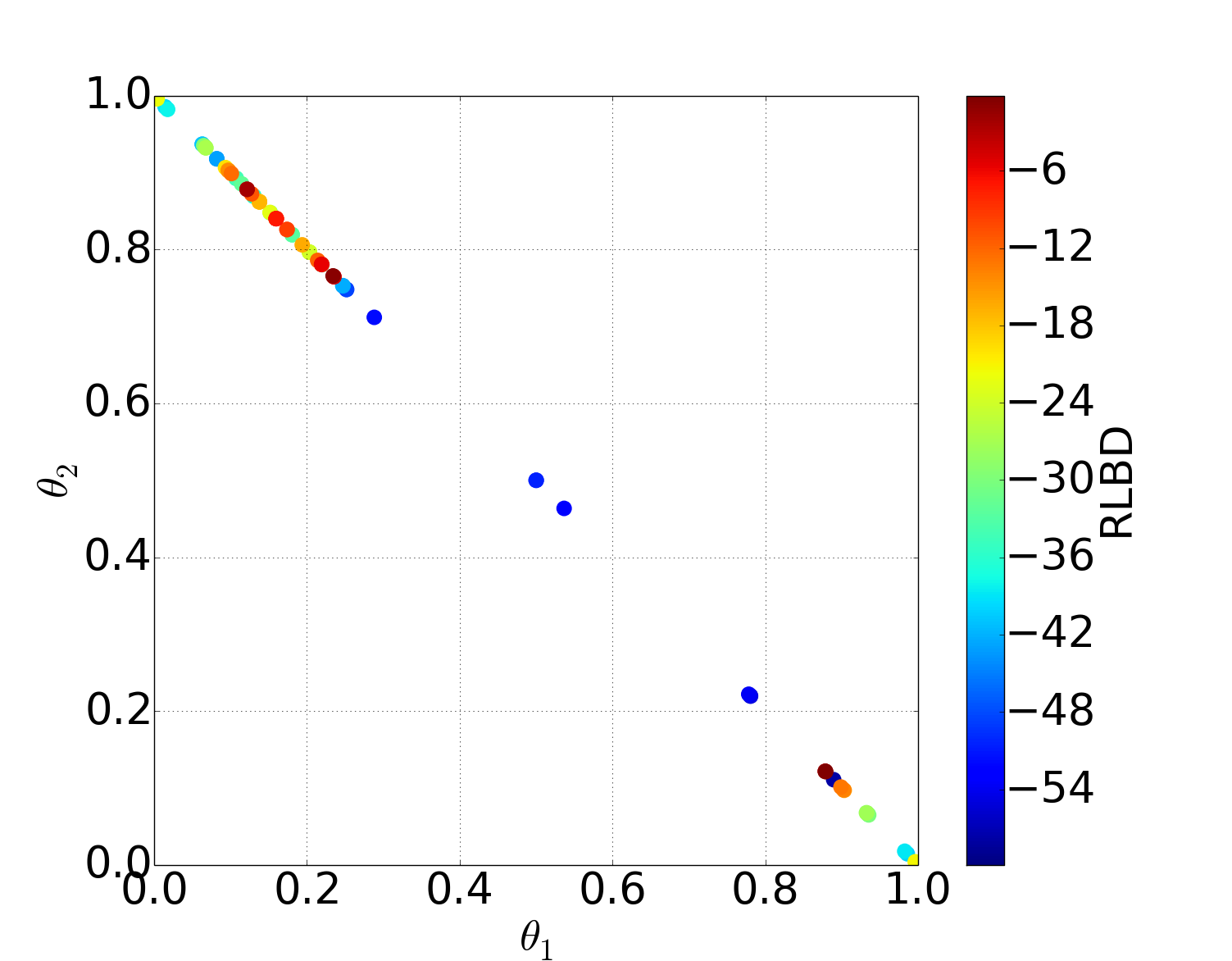}
\caption{Optimal $\theta$ of each iteration. 
The true $\theta$ is (0.22, 0.78).}
\end{subfigure}
\caption{Convergence and accuracy of our GOP algorithm on a synthetic data set on the scale of one feature, two subtypes, and ten samples.}
\label{fig:one_gene_ten_samples}
\end{figure}

We ran our GOP algorithm using 64 processors on a synthetic data set which is randomly generated on the scale of one feature ($M = 1$), two subtyes ($K = 2$) and ten samples ($N = 10$).
Figure~\ref{fig:one_gene_ten_samples} (a) shows that our GOP algorithm converges very quickly to -0.17 duality gap $(\PUBD - \RLBD)$ in the first 89 iterations in 120 seconds.
The optimal $x \ (x_1, x_2)$ and $\theta \ (\theta_1, \theta_2)$ of each iteration are shown with a range of colors to represent corresponding $\RLBD$ in Figure~\ref{fig:one_gene_ten_samples} (b, c). 
The dark blue represents low $\RLBD$ and the dark red represents high $\RLBD$. 
The $\RLBD$ of the initial $x$, $x^{(0)}$, is -59.87; The $\RLBD$ of iteration 89, $x^{(89)}$, is -0.17. 
It demonstrates that the GOP algorithm can change modes very easily without getting stuck in local optima. 

\subsection{Computational Complexity}
\label{subsec:computational_complexity}

We evaluate the GOP algorithm by theoretical analysis and empirical measurements of the time complexity on simulated data sets.
The problem has four main components: primal problem, preprocessing, unique region identification, and relaxed dual problems.  

\subsubsection{Theoretical Time Complexity}
\label{subsubsec:theoretical_time}

\paragraph{Primal problem}

The primal problem is a convex quadratic program with $KN$ decision variables. The time complexity for the primal problem solving is then $O(K^3N^3)$~\cite{boyd2004convex}.

\paragraph{Preprocessing}

We address the cases of overlapping qualifying constraint cuts by sorting the rows of the $KN \cdot M$ qualifying constraint coefficient matrix and comparing the coefficients of adjacent rows. 
We first sort the $KN$ rows of the qualifying constraint coefficient matrix using heapsort which takes $O(KN \cdot \text{log} (KN))$ time on average. 
The algorithm subsequently passes through the rows of the matrix to identify all-zero coefficients and duplicate cuts; each pass takes $O(KN)$ time.  
We define $|{\A^\prime}|$ as the number of unique qualifying constraints.

\paragraph{Unique region identification} 

The interior point method that we used in the A-star search algorithm is a linear program of size $|{\A^\prime}| \cdot MK$ with the time complexity of $O(|{\A^\prime}| \cdot MK)$. 
The time complexity for enumerating the set of unique regions is $O(|{\A^\prime}| \cdot (|{\A^\prime}| \cdot MK))$, which exhibits polynomial behavior.  
The time complexity of the partial order A-star algorithm is polynomial in the best case and exponential in the worst case, depending on the heuristic. 
We define $|r(\A^\prime)|$ as the number of identified unique regions.

\paragraph{Relaxed dual problems}

There are $2MK+1$ decision variables for each relaxed dual problem, so the time complexity for each is $O(M^3K^3)$. 
The total time for solving the relaxed dual problems is $O(|r(\A^\prime)| \cdot M^3K^3)$, which depends on the number of relaxed dual problems.


\subsubsection{Empirical Timing Results}
\label{subsubsec:empirical_time}

We constructed 12 synthetic data sets in a full-factorial arrangement with $M \in \left \{20, 40, 60, 80\right \}$, $K \in \left \{2\right \}$, and $N \in \left\{4, 5, 6 \right\}$ and measured CPU time for each component of one iteration.
For each arrangement, each element of the true $x^*$ is:

\vspace{5mm}
$x^*_{mk}=\left\{\begin{matrix}
1 & \text{if} \ 0\leqslant m < M/4, \ k = 0 \\ 
-1 & \ \text{if} \ M/4\leqslant m < M/2, \ k = 1\\
\mathcal{N}(0, 0.5^2) & \text{if} \ M/2\leqslant m < M, \forall k \\
0 & \text{otherwise}
\end{matrix}\right.$
\vspace{5mm}

Here $\mathcal{N}(0, 0.5^2)$ is the sample from a Normal distribution by its mean $0$ and standard deviation $0.5$.
For the true $\theta^*$, $\theta^*_{kn}$ for $k = 0$ are $n$ evenly spaced samples over the interval of $[0, 1]$;
$\theta^*_{kn}$ for $k = 1$ are $n$ evenly spaced samples over the interval of $[1, 0]$.


\begin{table}[htp]
  \centering
  \footnotesize
  \caption{Timing profile of each component of the GOP algorithm for one iteration.}
  \begin{threeparttable}
    \begin{tabular}{rrrrrrrr}
    \toprule
    \multicolumn{2}{c}{\textbf{Scale}} & \multicolumn{6}{c}{\textbf{Time (s)}} \\
    \cmidrule(r{4pt}){1-2} \cmidrule(l){3-8}
    \textbf{M}  & \textbf{N}  & \textbf{Primal} & \textbf{Pre} & \textbf{URI} & \textbf{Num} & \textbf{Dual} & \textbf{Total} \\ 
    \midrule
    20     & 4     & 0.10  & 1.69  & 1.29  & 200   & 1.54 (33\%)  & 4.62 \\
    40     & 4     & 0.12  & 1.91  & 1.72  & 202   & 1.69 (31\%)  & 5.44 \\
    60     & 4     & 0.12  & 2.03  & 1.11  & 202   & 1.77 (35\%)  & 5.03 \\
    80     & 4     & 0.13   & 2.39   & 2.05  & 232   & 3.70 (45\%)  & 8.27 \\
    \midrule
    20     & 5     & 0.11  & 1.99  & 1.31  & 456   & 11.26 (77\%)  & 14.67 \\
    40     & 5     & 0.11  & 2.07  & 1.37  & 485   & 11.45 (76\%) & 15.00 \\
    60     & 5     & 0.11  & 1.86  & 1.41  & 558   & 12.33 (78\%) & 15.71 \\
    80     & 5     & 0.12   & 2.23   & 1.26  & 650   & 17.96 (83\%) & 21.57 \\
    \midrule
    20     & 6     & 0.14  & 2.21  & 2.50  & 1152   & 65.71 (93\%) & 70.56 \\
    40     & 6     & 0.13  & 2.83 & 2.49  & 1250   & 67.08 (92\%) & 72.53 \\
    60     & 6     & 0.12  & 3.45  & 2.80  & 1255   & 69.00 (92\%) & 75.37 \\
    80     & 6     & 0.12  & 3.15  & 2.80  & 1309   & 77.62 (93\%) & 83.69 \\
    \bottomrule
    \end{tabular}
    \begin{tablenotes}
    \item Primal: primal problem. Pre: preprocessing. URI: unique region identification. Num: number of relaxed dual problems. Dual: relaxed dual problems. Total: total time of one iteration. Here we have two subtypes. A single processor is used. Time for solving relaxed dual problems is highlighted in percentage. 
    \end{tablenotes}
\end{threeparttable}
  \label{tab:profile}
\end{table}

Table \ref{tab:profile} shows that the time per iteration increases linearly with $M$ when $K$ and $N$ are fixed. 
The time for solving all the relaxed dual problems increases as the number of samples increases.
Even though the step of solving all the relaxed dual problems takes more than $90\%$ of the total time per iteration when the number of samples is $6$, our algorithm is easily parallelized to solve the relaxed dual problems, allowing the algorithm to scale nearly linearly with the size of the data set.


%
%

%
%
%

\section{Discussion}
\label{sec:discussion}

We have presented a global optimization algorithm for a mixed membership matrix factorization problem.
Our algorithm brings ideas from the global optimization community (Benders' decomposition and the GOP method) into contact with statistical inference problems for the first time.
The cost of the global optimal solution is the need to solve a number of linear programs that grows exponentially in the number of so-called ``connected'' variables in the worst case -- in this case the $KN$ elements of $\theta$.
Many of these linear programs are redundant or yield optimal solutions that are greater than the current upper bound and thus not useful. 
A branch-and-bound framework~\cite{Floudas1994} reduces the need to solve all possible relaxed dual problems by fathoming parts of the solution space
We further mitigate this cost by developing an search algorithm for identifying and enumerating the true number of unique linear programs.

We are exploring the connections between GOP and the other alternating optimization algorithms such as the expectation maximization (EM) and variational EM algorithm. 
Since the complexity of GOP only depends on the connected variables, the graphical model structure connecting the complicating and non-complicating variables may be used to identify the worst-case complexity of the algorithm prior to running the algorithm. 
A factorized graph structure may provide an approximate, but computationally efficient algorithm based on GOP. 
Additionally, because the Lagrangian function factorizes into the sum of Lagrangian functions for each sample in the data set, we may be able to update the parameters based on GOP for a selected subset of the data in an iterative or sequential algorithm. 
We are exploring the statistical consistency properties of such an update procedure.

Finally, we have derived an algorithm for particular loss functions for the sparsity constraint and  objective function. 
The GOP framework can handle integer variables and thus may be used with an $\ell_0$ counting ``norm'' rather than the $\ell_1$ norm to induce sparsity. 
This would give us a mixed-integer biconvex program, but the conditions for the framework. 
Structured sparsity constraints can also be defined as is done for elastic-net extensions of LASSO regression. 
It may be useful to consider other loss functions for the objective function depending on the application.

\subsubsection*{Acknowledgements}
We acknowledge Hachem Saddiki for valuable discussions and comments on the manuscript.

\bibliographystyle{plain}
\small
\bibliography{glad_gop}

\appendix
\onecolumn
\section*{SUPPLEMENTARY MATERIAL}

\section{Derivation of relaxed dual problem constraints}

The Lagrange function is the sum of the Lagrange functions for each sample,
\begin{equation}
L(y, \theta, x, \lambda) = \sum_{i=1}^n L(y_i, \theta_i, x, \lambda_i, \mu_i),
\end{equation}
and the Lagrange function for a single sample is
\begin{equation}
L(y_i, \theta_i, x, \lambda_i, \mu_i) = y_i^T y_i -2 y_i^T x\theta_i + \theta_i^T x^T x \theta_i - \lambda_i(\theta_i^T 1_K - 1) -\mu_i^T \theta_i.
\end{equation}

We see that the Lagrange function is biconvex in $x$ and $\theta_i$. We develop the constraints for a single sample for the remainder.

\subsection{Linearized Lagrange function with respect to $x$}

Casting $x$ as a vector and rewriting the Lagrange function gives
\begin{equation}
L(y_i, \theta_i, \bar{x}, \lambda_i, \mu_i) = a_i - 2b_i^T\bar{x} + \bar{x}^TC_i\bar{x} - \lambda_i(\theta_i^T 1_K - 1) -\mu_i^T\theta_i,
\end{equation}
where $\bar{x}$ is formed by stacking the columns of $x$ in order. The coefficients are formed such that
\begin{eqnarray*}
a                     &=&    y_i^T y_i, \\
b_i^T \bar{x}         &=& y_i^T x \theta_i, \\
\bar{x}^T C_i \bar{x} &=& \theta_i^T x^T x \theta_i.
\end{eqnarray*}

The linear coefficient matrix is the $KM \times 1$ vector,
\begin{equation*}
b_i = \left[ y_{i}\theta_{1i}, \cdots, y_{i}\theta_{Ki} \right]
\end{equation*}

The quadratic coefficient is the $KM \times KM$ and block matrix
\begin{equation*}
C_i = 
\left[
\begin{array}{ccc}
\theta^2_{1i} I_M & \cdots & \theta_{1i} \theta_{Ki} I_M \\
\vdots & \ddots & \vdots \\
\theta_{Ki} \theta_{1i} I_M & \cdots & \theta^2_{Ki} I_M 
\end{array} 
\right]
\end{equation*}

The Taylor series approximation about $x_0$ is
\begin{equation}
 L(y_i, \theta_i, \bar{x}, \lambda_i, \mu_i) \bigg|^{\text{lin}}_{\bar{x}_0} = L(y_i, x_0, \theta_i, \lambda_i, \mu_i) + (\nabla_x L \big|_{x_0})^T(x-x_0).
\end{equation}

The gradient with respect to $x$ is
\begin{equation}
\nabla_x L(y_i, \theta_i, \bar{x}, \lambda_i, \mu_i) = -2 b_i + 2 C_i \bar{x}.
\end{equation}

Plugging the gradient into the Taylor series approximation gives
\begin{equation}
L(y_i, \theta_i, \bar{x}, \lambda_i) \bigg|^{\text{lin}}_{\bar{x}_0} = a_i - 2b_i^T\bar{x}_0 + \bar{x}_0^TC_i\bar{x}_0 - \lambda_i(\theta_i^T 1_K - 1) - \mu_i^T \theta_i + (-2 b_i + 2 C_i \bar{x}_0)^T(\bar{x}-\bar{x}_0).
\end{equation}

Simplifying the linearized Lagrange function gives
\begin{equation}\label{eqn:lagrange_linx}
L(y_i, \theta_i, \bar{x}, \lambda_i, \mu_i) \bigg|^{\text{lin}}_{\bar{x}_0} = (y_i^T y_i - \bar{x}_0^T C_i \bar{x}_0 - \lambda_i(\theta_i^T 1_K - 1) - \mu_i^T \theta_i) - 2 b_i^T \bar{x} + 2 \bar{x}_0^T C_i \bar{x}
\end{equation}

Finally, we write the linearized Lagrangian using the matrix form of $x_0$,

\begin{equation}
L(y_i, \theta_i, x, \lambda_i, \mu_i) \bigg|^{\text{lin}}_{x_0} = 
y_i^T y_i^T - \theta_i^T x_0^T x_0 \theta_i - 2 y_i^T x \theta_i + 2 \theta_i^T x_0^T x \theta_i - \lambda_i(\theta_i^T 1_K - 1) - \mu_i^T \theta_i
\end{equation}

While the original Lagrange function is convex in $\theta_i$ for a fixed $x$, the linearized Lagrange function is not necessarily convex in $\theta_i$. This can be seen by collecting the quadratic, linear and constant terms with respect to $\theta_i$,

\begin{equation}
L(y_i, \theta_i, x, \lambda_i, \mu_i) \bigg|^{\text{lin}}_{x_0} = 
(y_i^T y_i^T + \lambda_i) + (- 2 y_i^T x -\lambda_i 1_K^T -\mu_i^T ) \theta_i + \theta_i^T (2 x_0^T x - x_0^T x_0 ) \theta_i.
\end{equation}

Now, if and only if $2x_0^Tx - x_0^Tx_0 \succeq 0$ is positive semidefinite, then $L(y_i, \theta_i, x, \lambda_i, \mu_i) \bigg|^{\text{lin}}_{x_0}$ is convex. The condition is satisfied at $x = x_0$ but may be violated at some other value of $x$.

\subsection{Linearized Lagrange function with respect to $\theta_i$}
Now, we linearize \eqref{eqn:lagrange_linx} with respect to $\theta_i$. Using the Taylor series approximation with respect to $\theta_{0i}$ gives

\begin{equation}
L(y_i, \theta_i, x, \lambda_i, \mu_i) \bigg|^{\text{lin}}_{x_0, \theta_{0i}} = L(y_i, \theta_{0i}, x, \lambda_i, \mu_i) \bigg|^{\text{lin}}_{x_0} + \left( \nabla_{\theta_i} L(y_i, \theta_i, x, \lambda_i, \mu_i) \bigg|^{\text{lin}}_{x_0} \bigg|_{\theta_{0i}} \right)^T (\theta_i - \theta_{0i})
\end{equation}

The gradient for this Taylor series approximation is
\begin{equation}
g_i(x) \triangleq \nabla_{\theta_i} L(y_i, \theta_i, x, \lambda_i, \mu_i) \bigg|^{\text{lin}}_{x_0} \bigg|_{\theta_{0i}} = -2 x_0^T x_0 \theta_{0i} -2 x^T y_i + 2 (x_0^T x + x^T x_0) \theta_{0i} - \lambda_i 1_K - \mu_i,
\end{equation}
where $g_i(x)$ is the vector of $K$ qualifying constraints associated with the Lagrange function. The qualifying constraint is linear in $x$.

Plugging the gradient into the approximation gives
\begin{equation}
\begin{split}
L(y_i, \theta_i, x, \lambda_i, \mu_i) \bigg|^{\text{lin}}_{x_0, \theta_{0i}} =\ & y_i^T y_i^T - \theta_{0i}^T x_0^T x_0 \theta_{0i} - 2 y_i^T x \theta_{0i} + 2 \theta_{0i}^T x_0^T x \theta_{0i} - \lambda_i(\theta_{0i}^T 1_K - 1) - \mu_i^T \theta_{0i} \\
& + (-2 x_0^T x_0 \theta_{0i} -2 x^T y_i + 2 (x_0^T x + x^T x_0) \theta_{0i} - \lambda_i 1_K - \mu_i)^T (\theta_i - \theta_{0i})
\end{split}
\end{equation}

The linearized Lagrange function is bi-linear in $x$ and $\theta_i$.

Finally, simplifying the linearized Lagrange function gives
\begin{equation}
\begin{split}
L(y_i, \theta_i, x, \lambda_i, \mu_i) \bigg|^{\text{lin}}_{x_0, \theta_{0i}} =\ & y_i^T y_i^T + \theta_{0i}^T x_0^T x_0 \theta_{0i} -2 \theta_{0i}^T x_0^T x_0 \theta_i - \lambda_i(1_K^T \theta_i - 1) - \mu_i^T \theta_i \\
& - 2 \theta_{0i}^T x^T x_0 \theta_{0i} - 2 y_i^T x \theta_i + 2 \theta_{0i}^T (x_0^T x + x^T x_0) \theta_i
\end{split}
\end{equation}

\section{Proof of Biconvexity} \label{bioconvex_proof}
To prove the optimization problem is biconvex, first we show the feasible region over which we are optimizing is biconvex.
Then, we show the objective function is biconvex by fixing $\theta$ and showing convexity with respect to $x$, and then vice versa.

\subsection{The constraints form a biconvex feasible region}
Our constraints can be written as
\begin{align}
||x||_1 & \leqslant P \label{eqn:1}\\
\sum_{k=1}^{K}\theta_{ki} & = 1 \  \forall i \label{eqn:2} \\
0 \leqslant \theta_{ki} & \leqslant 1  \  \forall (k, i). \label{eqn:3}
\end{align}

The inequality constraint \eqref{eqn:1} is convex if either $x$ or $\theta$ is fixed, because any norm is convex.
The equality constraints \eqref{eqn:2} is an affine combination that is still affine if either $x$ or $\theta$ is fixed.
Every affine set is convex.
The inequality constraint \eqref{eqn:3} is convex if either $x$ or $\theta$ is fixed, because $\theta$ is a linear function.

\subsection{The objective is convex with respect to $\theta$}

We prove the objective is a biconvex function using the following two theorems.

\begin{theorem}
Let $A \subseteq {\mathbb{R}^n}$ be a convex open set and let $f: A \rightarrow \mathbb{R}$ be twice differentiable. Write $H(x)$ for the Hessian matrix of $f$ at $x\in A$. If $H(x)$ is positive semidefinite for all $x\in A$, then $f$ is convex (\cite{boyd2004convex}).
\end{theorem}

\begin{theorem}
$A$ symmetric matrix $A$ is positive semidefinite (PSD) if and only if there exists $B$ such that $A = B^TB$ (\cite{lancaster1985theory}).
\end{theorem}

The objective of our problem is,
\begin{align}
\label{eq:test}
f(y,x,\theta) = ||y-x \theta||^2_2 & = (y-x\theta)^T(y-x\theta) \\
& = (y^T-\theta^Tx^T)(y-x\theta)\\
& = y^Ty - y^Tx\theta - \theta^Tx^Ty + \theta^Tx^Tx\theta.
\end{align}

The objective function is the sum of the objective functions for each sample.
\begin{align}
f(y,x,\theta) & =\sum_{i=1}^{N}f(y_i,x,\theta_i)\\
& = \sum_{i=1}^{N}y_i^T y_i -2y_i^T x \theta_i + \theta_i^T x^T x \theta_i.
\end{align}

The gradient with respect to $\theta_i$,
\begin{align}
\nabla _{\theta_i}f(y_i,x,\theta_i)&= -2 y_i^T x+ (x^Tx+(x^Tx)^T)\theta_i\\
& = -2 x^Ty_i + 2x^Tx\theta_i.
\end{align}
Take second derivative with respect to $\theta_i$ to get Hessian matrix,
\begin{align}
\nabla _{\theta_i}^2 f(y_i,x,\theta_i)&= \triangledown _{\theta_i}( -2 x^Ty_i + 2x^Tx\theta_i )\\
& = 2 \triangledown _{\theta_i} (x^Tx\theta_i)\\
& = 2 (x^Tx)^T\\
& = 2 x^Tx.
\end{align}
The Hessian matrix $\nabla_{\theta_i}^2 f(y_i,x,\theta_i)$ is positive semidefinite based on Theorem 2.
Then, we have $f(y_i,x,\theta_i)$ is convex in $\theta_i$ based on Theorem 1.
The objective $f(y,x,\theta)$ is convex with respect to $\theta$, because the sum of convex functions, $\sum_{i=1}^{N}f(y_i,x,\theta_i)$, is still a convex function.

\subsection{The objective is convex with respect to $x$}
The objective function for sample $i$ is
\begin{align}
f(y_i, x, \theta_i) = y_i^Ty_i - 2y_i^Tx\theta_i + \theta_i^Tx^Tx\theta_i.
\end{align}
We cast $x$ as a vector $\bar{x}$, which is formed by stacking the columns of $x$ in order.
We rewrite the objective function as
\begin{align}
f(y_i, \bar{x}, \theta_i)=a_i - 2b_i^T\bar{x} + \bar{x}^TC_i\bar{x}.
\end{align}
The coefficients are formed such that
\begin{align}
a & = y_i^Ty_i,\\
b_i^T\bar{x} &= y_i^Tx\theta_i,\\
\bar{x}^TC_i\bar{x} &=\theta_i^Tx^Tx\theta_i.
\end{align}
The linear coefficient matrix is the $KM \times 1$ vector
\begin{align}
b_i=[y_i\theta_{1i},...,y_i\theta_{Ki}]
\end{align}
The quadratic coefficient is the $KM \times KM$ and block matrix
\begin{align}
C_i=\begin{bmatrix}
\quad\theta_{1i}^2I_M \quad\cdots\quad \theta_{1i}\theta_{Ki}I_M & \\
\vdots \qquad \ddots \qquad \vdots & \\
\theta_{Ki}\theta_{1i}I_M\quad \cdots\quad \theta_{Ki}^2I_M&
\end{bmatrix}
\end{align}
The gradient with respect to $\bar{x}$
\begin{align}
\nabla_{\bar{x}}f(y_i,\bar{x},\theta_i)& = -2b_i + 2C_i\bar{x}.
\end{align}
Take second derivative to get Hessian matrix,
\begin{align}
\nabla_{\bar{x}^2}f(y_i,\bar{x},\theta_i)& = 2C_i^T\\
& = 2(\theta_i\theta_i^T)^T\\
& = 2(\theta_i^T)^T(\theta_i^T).
\end{align}
The Hessian matrix $\nabla _{\bar{x}}^2 f(y_i,\bar{x},\theta_i)$ is positive semidefinite based on Theorem 2.
Then, we have $f(y_i,\bar{x},\theta_i)$ is convex in $\bar{x}$ based on Theorem 1.
The objective $f(y,x,\theta)$ is convex with respect to $x$, because the sum of convex functions, $\sum_{i=1}^{N}f(y_i,x,\theta_i)$, is still a convex function.

The objective is biconvex with respect to both $x$ and $\theta$.
Thus, we have a biconvex optimization problem based on the proof of biconvexity of the constraints and the objective.

\section{A-star Search Algorithm} \label{A}
In this procedure, first we remove all the duplicate and all-zero coefficients hyperplanes to get unique hyperplanes.
Then we start from a specific region $r$ and put it into a open set.
Open set is used to maintain a region list which need to be explored.
Each time we pick one region from the open set to find adjacent regions.
Once finishing the step of finding adjacent regions, region $r$ will be moved into a closed set.
Closed set is used to maintain a region list which already be explored.
Also, if the adjacent region is a newly found one, it also need to be put into the open set for exploring.
Finally, once the open set is empty, regions in the closed set are all the unique regions, and the number of the unique regions is the length of the closed set.
This procedure begins from one region and expands to all the neighbors until no new neighbor is existed.

The overview of the A-star search algorithm to identify unique regions is shown in Algorithm 2.
\begin{algorithm}
\caption{A-star Search Algorithm}
\begin{algorithmic}[1]
\STATE Sort the rows of the $KN$ x $M$ qualifying constraint coefficient matrix.
\STATE Compare adjacent rows of the qualifying constraint coefficient matrix and eliminate duplicate rows.
\STATE Eliminate rows of the qualifying constraint coefficient matrix with all-zero coefficients.
\STATE Determine the list of unique qualifying constraints by pairwise test.
\STATE Set $S$ and $|\A^\prime|$ to the set of unique, non-trivial qualifying constraints and the number of them.
\STATE Initialize a region $root$ using an interior point method (Component 1).
\STATE Put region $root$ into the open set.
\IF {open set is not empty}
  \STATE Get a region $R$ from the open set.
  \STATE Calculate the adjacent regions set $R\_adj$ (Component 2).
  \STATE Put region $R$ into the closed set.
  \FOR {each region $r$ in $R\_adj$}
    \IF {$r$ is not in the open set $and$ not in the closed set}
      \STATE Put region $r$ into the open set.
    \ENDIF
  \ENDFOR
\ENDIF
\STATE Reflect the sign of the regions in the close set.
\STATE Get all the regions represented by string of 0 and 1.
\end{algorithmic}
\end{algorithm}

\paragraph{Hyperplane filtering} \label{duplicate}
Assuming there are two different hyperplanes $H_i$ and $H_j$ represented by $A_i=\left\{a_{i,0},...,a_{i,MK}\right\}$ and $A_j=\left\{a_{j,0},...,a_{j,MK}\right\}$.
We take these two hyperplanes duplicated when
\begin{equation}
	 \frac{a_{i,0}}{a_{j,0}}=\frac{a_{i,1}}{a_{j,1}}=...=\frac{a_{i,MK}}{a_{j,MK}}=\frac{\sum_{l=0}^{MK}a_{i,l}}{\sum_{l=0}^{MK}a_{j,l}}, a_{j,l}!=0
\end{equation}
This can be converted to
\begin{equation}
	|\sum_{l=0}^{MK}a_{i,l}\cdot a_{j,n}-\sum_{l=0}^{MK}a_{j,l}\cdot a_{i,n}|\leqslant \tau, \forall \ n \epsilon [0,MK]
\end{equation}
where threshold $\tau$ is a very small positive value.

We eliminate a hyperplane $H_i$ represented by $A_i=\left\{a_{i,0},...,a_{i,MK}\right\}$ from hyperplane arrangement $\A$ if the coefficients of $A_i$ are all zero,
\begin{flalign}
	|a_{i,j}|\leqslant \tau,
	&\ \forall \ a_{i,j} \epsilon A_i,
	j\epsilon [0,MK]
\end{flalign}
$\A^\prime$ is the reduced arrangement and $A^\prime x=b$ are the equations of unique hyperplanes.

\paragraph{Interior point method}
An interior point is found by solving the following optimization problem:
\begin{flalign}
\text{maximize} &\  z \nonumber\\
\text{subject to} &\  -A^\prime_ix + z \leqslant b_i, \text{if}\ \theta^B_i =0\\
		  &\  A^\prime_ix + z \leqslant -b_i, \text{if}\ \theta^B_i =1\\
&\    z>0
\end{flalign}

\begin{algorithm}
\caption{Interior Point Method (Component 1)}
\begin{algorithmic}[1]
\STATE Generate $2^{|\A^\prime|}$ different strings using $0$ and $1$.
  \FOR {each $s$ in the strings}
    \STATE Solve an optimization problem to get an interior point.
    \IF {Get a interior point}
    \STATE Get the $root$ region represented by $0$ and $1$.
    \ENDIF
  \ENDFOR
\end{algorithmic}
\end{algorithm}

\begin{algorithm}
\caption{Get Adjacent Regions (Component 2)}
\begin{algorithmic}[1]

\STATE Initialize an empty set $SH$ for strict hyperplanes.
\STATE Initialize an adjacent region set $ADJ$.

\STATE \# Find out all the strict hyperplanes for region $R$.
\FOR {each hyperplane $H$ of $|\A^\prime|$ hyperplanes}
 \STATE Pick one hyperplane $H$ from all the hyperplanes defining region R.
 \STATE Flip the sign of $H$ to get $\neg H$.
 \STATE Form a new hyperplane arrangement $\neg \A^\prime$ with $\neg H$.
 \STATE Solve the problem to get an interior point constrained by $\neg \A^\prime$.
 \IF {the interior point is not Non}
 \STATE $H$ is a strict hyperplane and put into set $SH$.
 \ELSE
 \STATE $H$ is a redundant hyperplane.
 \ENDIF
\ENDFOR

\STATE \# Find out all the adjacent regions for region $R$.
\FOR {each strict hyperplane $sh$ in set $SH$}
 \STATE Take the opposite sign $\neg sh$ of $sh$.
 \STATE Form a adjacent region $adj$ based on $\neg sh$ and all the else hyperplanes.
 \STATE Put $adj$ into set $ADJ$.
\ENDFOR

\end{algorithmic}
\end{algorithm}

\newpage

\end{document}